\documentclass[a4paper,11pt]{article}
\pdfoutput=1 

\usepackage{jcappub} 

\usepackage[T1]{fontenc} 

\usepackage{graphics}
\usepackage{array}
\usepackage{stfloats}
\usepackage{fixltx2e}
\usepackage{amsmath}
\usepackage{fontenc}
\usepackage{amssymb,amsfonts,txfonts}
\usepackage{threeparttable}
\usepackage{longtable}
\usepackage{graphicx}
\usepackage{url}
\usepackage{textcomp}
\usepackage{color}
\usepackage{times}
\usepackage[subnum]{cases}
\fontencoding{T1}

\usepackage{bm}

\newcommand{\be}{\begin{equation}}
\newcommand{\ee}{\end{equation}}
\newcommand{\bea}{\begin{eqnarray}}
\newcommand{\eea}{\end{eqnarray}}

\def\ltsima{$\; \buildrel < \over \sim \;$}
\def\simlt{\lower.5ex\hbox{\ltsima}}
\def\gtsima{$\; \buildrel > \over \sim \;$}
\def\simgt{\lower.5ex\hbox{\gtsima}}

\def\calR{\mathcal{R} }
\def\calC{\mathcal{C} }

\def\T{\mathbb{T}}
\def\arr{\rightarrow}

\def\si{\sigma}
\def\ga{\gamma}
\def\ka{\kappa}
\def\de{\delta}
\def\na{\nabla}
\def\Om{\Omega}
\def\al{\alpha}

\def\ep{\epsilon}
\def\({\left(}
\def\){\right)}

\def\om{\omega}
\def\Ga{\Gamma}

\title{A further study on Palatini $f(\calR)$-theories for polytropic stars} 

\author[a,b]{Annalisa Mana}
\author[c,d]{, Lorenzo Fatibene}
\author[c]{and Marco Ferraris}

\affiliation[a]{Universit\"ats-Sternwarte, Ludwig-Maximilians Universit\"at, Scheinerstr. 1, 81679 Munich, Germany}
\affiliation[b]{Excellence Cluster Universe, Boltzmannstr. 2, 85748 Garching, Germany}
\affiliation[c]{Department of Mathematics, University of Torino, Italy}
\affiliation[d]{INFN Sezione Torino, Iniz. Spec. QGSKY, Italy}

\emailAdd{amana@usm.lmu.de, lorenzo.fatibene@unito.it, marco.ferraris@unito.it}

\abstract{
After briefly reviewing the results about polytropic stars in Palatini $f(\calR)$-theories, we first show how these results rely on the assumption of a regular function $f(\calR)$. In particular, singular models allow to extend the parameter interval in which no singularity is formed.

Furthermore,  we show how the conformal metric can be matched smoothly in the cases where the original metric generates a singularity.
In fact, the singularity comes from a singular conformal factor which is continuous though not  differentiable at the stellar surface.
This suggests that the correct metric to be considered as physical is the conformal metric.

This is relevant because, even when matching the original metric is possible, the use of the conformal metric  generates different stellar models.
}

\arxivnumber{1505.06575}

\notoc

\begin{document}
\maketitle

\section{Introduction}

Stellar models with polytropic equation of state (EoS) in Palatini $f(\calR)$-theories generically lead  to a singularity which forms near the stellar surface for some parameters of the EoS \citep[see][]{no-go}.
Since no extreme physical situation is expected near the surface, this  result has been used to rule out Palatini $f(\calR)$-theories.

A further analysis \citep[see][]{Olmo} relates the singularity to the particular form of EoS, rather than to the dynamics of $f(\calR)$.
In particular, it is shown that the singularity forms at pressures which are unphysically low. It is thus argued that the singularity can be removed
by  a slight modification of the EoS. 
Since the EoS is always an approximation of the matter equations in a selected dynamical regime, a mathematical singularity which is unstable under modifications of the EoS needs a detailed discussion of the EoS itself, in order to be considered on a physical stance.

Hereafter, we investigate the assumptions at the basis of the no-go theorem for polytropic stars in Palatini $f(\calR)$-theories.
In particular, we show that allowing continuous non-analytical $f(\calR)$ dynamics, one can extend the parameter interval in which no singularity forms to cover the physical cases (for example, neutron stars), which were previously left out.
The regularity assumption on the function $f(\calR)$ hence appears to be essential for the no-go theorem.

Palatini $f(\calR)$-theories are essentially bimetric theories with an {\it original metric} $g$ and a {\it conformal metric} $\tilde g$.
Previous analyses were done as if the conformal metric is an auxiliary one, while the original metric is endowed with the whole physical meaning. 
 
We have different reasons to consider alternatives.
In the beginning of the '70s Ehlers-Pirani-Schild (EPS) proposed an axiomatic approach to gravitational theories, in which the geometry of spacetime is derived from potentially observable quantities, instead of being assumed {\it a priori} \cite[see][]{EPS,EPS1,EPS2}).
EPS framework clearly singles out a conformal structure, which is defined by light rays, and a free fall related projective structure.
These two structures have to be compatible, in order to implement light cones as limit speed for free fall: 
this condition is called {\it EPS-compatibility}.  When this holds the geometry of spacetime is a Weyl geometry, not necessarily a Lorentzian structure.

Thus, it is natural to consider Palatini theories for a metric and a connection: the first related to light cones and distances, the second related to free fall.
In addition, we need to restrict to those dynamics which implement the EPS-compatibility between conformal and projective structure.
These theories are called {\it extended theories of gravitation} (ETG) and Palatini $f(\calR)$-theories are in fact ETG.

It can then be assumed that the connection $\tilde \Ga$, and not the original metric, is related to free fall. By solving the equations and introducing the conformal metric $\tilde g$, the latter  is thus endowed with the physical meaning of determining free fall of test particles.
We  show that in stellar models matching conformal metric is as easy as in standard GR and singularities at the surface do not generically form.
It is in view of this result that one can consider the conformal metric as a candidate for the physical metric, even ignoring EPS framework.

In conclusion, the matching of the conformal metric clearly spots the origin of surface singularities in the derivative of the conformal factor.
As former analyses of no-go theorem demonstrated, these singularities are generic with respect to the function $f(\calR)$.
Moreover,  matching the original or the conformal metric is always different, even for the parameter values which allow regular matching
in both cases.

\subsection{No-go theorem}
\label{nogo_theorem}
In \cite[][]{no-go} it was shown that, for generic choices of analytical $f(\calR)$, there are commonly used equations of state for which no satisfactory physical solution of the field equations can be found (apart from the special case of GR), casting doubt on whether Palatini $f(\calR)$ gravity can be considered as a physically sound alternative to GR.


We consider the common 
action for Palatini theories on a spacetime $\mathcal{M}$ of dimension $m=\dim(\mathcal{M})$ given as
\be
\label{eq:1}
S[g, \widetilde{\Gamma},\psi]=\frac{1}{2{\kappa}^2}\int \mathrm{d}^{m}x \sqrt{-g} \left[f(\calR)+\mathcal{L}_{\mathcal{M}}(g,\psi)\right]\ ,
\ee
where we use relativistic units, i.e. $c = \hbar = 1$. Here, ${\kappa}^{2} \equiv 8\pi G$, $G$ being the gravitational constant, $g$ is the determinant of the metric $g_{\mu\nu}$ of signature $(m-1,1)$, $\calR \, [g,\widetilde{\Gamma}]\, = g^{\mu\nu}\widetilde{R}_{\mu\nu}$ is the Ricci scalar built 
with an independent torsion-less connection $\tilde \Gamma$, $\psi$ denotes a collection of matter fields, $\mathcal{L}_{M}$ is the matter Lagrangian (here assumed to be independent of $\tilde \Gamma$) and $f(\calR)$ is an analytic function of the scalar curvature.

The matter Lagrangian is assumed not to depend on the connection $\tilde\Gamma$. 
This is a standard assumption in Palatini $f(\calR)$ gravity and it makes easier to discuss field equations hereafter.
If this assumption is relaxed one has an extra stress tensor on the right hand side of the second field equation (\ref{eq:2}) below.
These models have been discussed in the literature;  \cite[see][]{S1,S2, MatterCouplings}.
However, generically, this extra coupling prevents the connection $\tilde\Gamma$ from being metric and the spacetime to be endowed with an integrable Weyl geometry;  \cite[see][]{ExtendedTheories}. These more general models can be studied even if they are affected by holonomy problems as the original Weyl proposal for unified theories of gravity and electromagnetism. 

Independent variations with respect to the metric $g_{\mu\nu}$ and the connection $\tilde \Gamma$ give the following field equations
\begin{numcases}
 \, f'(\calR)\,\widetilde{R}_{(\mu\nu)}-\frac{1}{2}f(\calR)g_{\mu\nu} = {\kappa}^{2} T_{\mu\nu}\ , \\
\widetilde{\nabla}_{\lambda}\left[\sqrt{-g}f'(\calR)g^{\mu\nu}\right] = 0\ ,
\label{eq:2}
\end{numcases}
where $f'(\calR)\equiv \mathrm{d}f(\calR) / \mathrm{d}\calR $, $\widetilde{\nabla}_{\mu}$ is the covariant derivative with respect to the independent connection $\widetilde \Gamma$ and $T_{\mu\nu}$ is the usual matter stress-energy tensor. 
We assume the latter in the standard form of a perfect fluid stress tensor, 
\be
\label{eq:tmunu}
T_{\mu\nu} = (\rho + p)u_{\mu}u_{\nu}+pg_{\mu\nu}\ ,
\ee
where $u$ is the $m$-velocity, $\rho$ the energy density and $p$ the pressure of the fluid. 
The length of  $u$ has to be fixed so that $|u|^2=-1$.

Equation (1.2a) can be traced by $g^{\mu\nu}$ to obtain the so-called {\it master equation}
\be
f'(\calR)\,\calR-\frac{m}{2}f(\calR)={\kappa}^{2} T\ ,
\label{eq:3}
\ee
where we set $T \equiv g^{\mu\nu} T_{\mu\nu}$. For a generic algebraic function $f$ the master equation is as well an algebraic equation for $\calR$, with $T$ considered as a parameter, and one can solve equation (\ref{eq:3}) for $\calR$ to obtain $\calR\, {\equiv}\, \calR(T)$. 
With the latter, we can redefine $f\, \equiv \, f[\calR(T)]$ and analogously for its derivatives. 
The derivative of $\calR$ with respect to $T$ is given as
\be
\calR_T \, \equiv \, \frac{\mathrm{d}\calR(T)}{\mathrm{d}T}= \frac{\kappa^2}{ f''\calR(T)+ \(1- \frac{m}{2}\) f'} \ .
\ee
Equation (1.2b) can be solved by defining a new conformal metric $\tilde g_{\mu\nu}= \phi \, g_{\mu\nu}$, with a conformal factor $\phi= (f')^{\frac{2}{m-2}}$. Consequently it becomes
\be 
\widetilde{\nabla}_{\lambda}\left[\sqrt{-\tilde g}\tilde g^{\mu\nu}\right]=0\ 
\label{eq:widetildenabla}
\ee
and can be solved with respect to the
connection to obtain 
\be
\widetilde\Gamma^{\lambda}_{\ \mu\nu}= \{\tilde g\}^{\lambda}_{\ \mu\nu}\ .
\label{eq:h}
\ee

In the end, by substituting equation (\ref{eq:3}) and the connection (\ref{eq:h}) into the first field equation (1.2a), one gets
\be
\label {eq:4}
\widetilde G_{\mu\nu}=\frac{{\kappa}^{2}}{f'}T_{\mu\nu}-\frac{1}{2}g_{\mu\nu}\left(\calR-\frac{f}{f'}\right) \, \equiv\, \kappa^2 \widetilde \T_{\mu\nu} \ ,
\ee 
where $\widetilde G_{\mu\nu}{\equiv} \tilde R_{\mu\nu}-\frac{1}{2} \tilde R \tilde g_{\mu\nu}$ is the Einstein tensor for the conformal metric $\tilde g$. Hereafter, tildas refer to quantities depending on the conformal metric $\tilde g$.

In view of conformal equivalence, one can decide to write field equations in terms of $\tilde g$ or $g$.
The field equation (\ref{eq:4}) written in terms of $g$ reads as 
\be
G_{\mu\nu}=\ka^2\T_{\mu\nu}\ ,
\label {fe}
\ee 
where we set
\be
\kappa^2\T_{\mu\nu} \equiv \frac{{\kappa}^{2}}{f'} T_{\mu\nu} - \frac{1}{2}\left(\calR - \frac{f}{f'}\right)g_{\mu\nu} 
+\frac{1}{f'} \left(\nabla_{\mu} \nabla_{\nu} f'-  \Box f' g_{\mu\nu}\right) -  \frac{m-1}{m-2}\frac{1}{(f')^2}\left(\nabla_{\mu}f' \nabla_{\nu}f'
-  \frac{1}{2 }   \nabla_{\rho} f' \nabla^{\rho} f' g_{\mu\nu}\right)\ .
\label {eq:4bis}
\ee 
Here, $\nabla_{\mu}$ is the covariant derivative with respect to $g$ and $\Box \equiv g^{\mu\nu}\nabla_{\mu}\nabla_{\nu}$ is the box operator of $g$. Setting the dimension $m=4$, one recovers the field equations in \cite{no-go}.

It is well known that Palatini $f(\calR)$-models in vacuum are equivalent to standard Einstein models with a cosmological constant dictated by the master equation \citep[see][]{Universality}. Hence, vacuum solutions are 
defined and we can assume exterior solution to be determined. To calculate the inner solution, we have to solve
(\ref{fe}) by assuming a static spherically symmetric ansatz 
\be\label{InSol}
g\,=\,-A(r)\,\mathrm{d}t^{2}\,+\,B(r)\,\mathrm{d}r^{2}\,+\,r^{2}\,\mathrm{d}\Omega^{2}\ ,
\ee
where $\mathrm{d}\Omega^{2}$ is the line element along the $S^{m-2}$ angular directions.
The $m$-velocity is then $u= \sqrt{A}/A\,\partial_0$.

We present the results of \cite{no-go} for dimension $m=4$ in the Appendix. {We checked that the result is similar in dimension $m=3$.}

Furthermore, if one assumes a polytropic Equation of State (pEoS) and tries to match the inner and exterior (vacuum) solutions, there are values of the polytropic parameters for which singularities develop on the surface of the star, namely there is no $\calC^1$ matching. This result holds for generic analytical functions $f(\calR)$ and for physically sound values of the polytropic parameters.

\subsection{Re-examination of polytropic star models}
\label{re-exam}

In \cite{Olmo}, the situation described in Section \ref{nogo_theorem} was reviewed in the particular case of
\be
\label{eq:f_r}
f(\calR) = \calR + \frac{\ep}{2} \calR^2	\ .
\ee
There is a natural scale for the constant $\epsilon$ which is of order the squared Planck length, for dimensional reasons.

Under these assumptions,  we can compute the divergent contributions to 
the scalar curvature $R$ of the metric $g$, as $p$ tends to $0$, namely
\be
R\sim \frac{(2-\ga)\ka^2\ep}{(\ga K)^2} \frac{1}{3\left(1-\frac{2M_0}{r_0}\right)} \>\rho^{3-2\ga}+\calR_0 \ .
\ee 
We refer to the Appendix for details and notation. Here in particular,   $\calR_0$ denotes the outer scalar curvature, $M_0$ and $r_0$
the mass and radius of the star, $K$ and $\ga$ are the parameters of the polytropic equation of state (pEoS); see equation (\ref{eq:EoS}) in the Appendix. 

The scalar curvature $R$, in fact, diverges as $\rho$ tends to $0$, for $3/2 < \ga < 2$. 
In common astrophysical models, for example neutron stars for which typically 
$\ga=5/3$, the divergence is driven by the a-dimensional factor and defines a critical density scale at which the singularity is forming, namely
\be
 \rho_c \simeq\( \frac{ K^2}{\ka^2\ep}\)^{\frac{1}{3-2\ga}} \ .
\ee 
For realistic
neutron stars one obtains $\rho_c \simeq 10^{-210}\, [\mathrm{g}/\mathrm{cm}^{3}]$, which is well below any reasonable physical density. 
Therefore, the existence of a curvature singularity at such extremely low densities should be considered as unphysical and an artifact of the idealized situation. To have an idea of this density scale, one single electron in the volume of a sphere with the radius of the Neptune orbit amounts to a density of $\rho_e\simeq  10^{-72} \, [\mathrm{g}/\mathrm{cm}^{3}]$.

In conclusion, the results of \cite{Olmo} have confirmed that the surface singularity exists, but its physical feature
is conjectured to depend on the details of the model. In Planck scale models, the singularity develops in conditions well beyond the validity of assumptions and approximations of the model itself.

\subsection{Problems and contents}
\label{contents}


Generally, Palatini $f(\calR)$ models together with pEoS predict a singularity at very low pressure (or density), for some intervals of polytropic parameter $\ga$. Although this prevents us from modelling some types of stars as usually done in standard GR, one has to consider that pEoS is not a fundamental principle of physics, being instead an approximation. 
For example, one could 
assume some mechanism which modifies the EoS near the surface preventing the formation of the singularity. 
Alternatively, one could relax the assumption of a vacuum solution in the exterior, as we do in the following.

Another interesting issue is related to the choice of the conformal frame. 
Both the primary no-go theorem analysis \citep[see][]{no-go} and its re-examination \citep[see][]{Olmo} are concerned with the original metric $g$, entering the variational principle. We know,  however that another natural metric $\tilde g$ arises while solving the field equations.  
In the case of a non-regular $f$ (or multiple zeros), the new metric $\tilde g$ can gain/lose regularity, by means of 
the conformal factor.\\

This paper is organized as follows. In Section \ref{sec:non-analytic-models}, we discuss the results revised in the Appendix. We show that if one relaxes the assumptions about the analyticity of the function $f$, then the quantity driving the singularity development may improve --quite drastically-- its behaviour. Ideally, after showing that the proof of the no-go theorem is not valid in these models, one should propose a model in which singularities can be avoided in the parameter range $\ga \in (0,2)$. Future investigations will be devoted to this topic. Section \ref{sec:cheese} examines the case of a polytropic star by embedding the spherical inner object into a Friedmann-Lema\^{i}tre-Robertson-Walker (FLRW) background: the so-called \emph{Swiss-cheese} model. However, the fact that singularities form at an unphysically low density is not enough to casually relate low pressures and singularities. Since it is difficult to obtain general results on the matching, one should be able to give an example, avoiding low pressures, in which the singularity does not appear. 
This turns out to be non-trivial, though investigating in this direction clarifies that the matching on a hypersurface has to be performed using Darmois junction conditions, instead of matching the metric coefficients only \citep[see][]{Dyer}. Darmois conditions are powerful because we can avoid assuming a coordinate system around the matching surface. In addition to this, one does not have to assume that the chart is precisely the one used to express the ansatze. More specifically, Darmois junction conditions require that the first fundamental forms (intrinsic metrics) and the second fundamental forms (extrinsic curvatures), calculated in terms of the respective coordinates (outside and inside the matching surface), must be identical on both sides. These junction conditions allow us to use different coordinate systems on both sides of the matching hypersurface. 
In Section \ref{sec:conf_trans}, we discuss how the matching procedure behaves under conformal transformations. This is motivated by the fact that the model is known to be similar to standard GR, with respect to $\tilde g$, and therefore being the matching possible. We find that the singularities are driven by a discontinuity in the derivative of the conformal factor. In the end, we draw our conclusions and physical interpretation in Section \ref{sec:concl}.


\section{Non-analytic models}
\label{sec:non-analytic-models}
We refer hereafter to the Appendix for the detailed calculations and notation. 
There  we review the original argument and find that one can define a quantity 
\be
C= (\rho+p) \frac{df'}{dp}
\ee
Here $f'$ denotes the derivative of the function $f(\calR)$ with respect to its argument $\calR$ then evaluated at $\calR(T)$ with $T=3p-\rho$.
Accordingly, it is a function of $(p,\rho)$. By using EoS one has $\rho=\rho(p)$ and $C$ becomes a function of $p$ only,
which is in turn a function $p(r)$ of the coordinate $r$ when evaluated along a solution.
Let us then denote by $C_r$ the derivative
\be
C_r= \frac{d C}{dp} \frac{d p}{dr} 
\ee

We focus now on the relevant quantity which drives the divergence of the scalar curvature of $g$, see equation (\ref{DivergentR}) in the Appendix, {namely} $C_r/f'$. Note that if one allows $f$ to be non-analytic, then $f'$ might diverge, implying consequently the convergence of the scalar curvature. 

For example, 
if we consider models with $f(\calR)= c_1 \calR^\nu$, then the derivative with respect to $\calR$ is given by $f'= \nu c_1 \calR^{\nu-1}$ and the master equation ({for $m=4$}) reads as
\be
f'\calR-2f= c_1(\nu-2) \calR^\nu=T\ .
\ee
From the latter, we obtain: 
\be
\calR = \left[ \frac{T}{c_1(\nu-2)}\right] ^{1/\nu}\ ,\ \ \ \ \ f'=\frac{ \nu c_1^{1/\nu}}{ (\nu-2)^{\frac{\nu-1}{\nu}} } T^{\frac{\nu-1}{\nu}}\ .
\ee 
By assuming a pEoS, we can infer the following behaviours of $f'$ with respect to $p$: 
\be
\label{eq:f_prime_cases}
{f' \sim}
\begin{cases}
p ^{\frac{\nu-1}{\nu}} & \mbox{for}\ \ 0<\ga<1\ ,\\
p ^{\frac{\nu-1}{\ga \nu}} & \mbox{for}\ \ 1<\ga<2\ .
\end{cases}
\ee
From equation (\ref{eq:f_prime_cases}), it follows that
\be
\frac{C_r}{f'} \sim 
\begin{cases}
 p ^{\frac{1}{\nu}}& \mbox{for}\ \ 0<\ga<1\ ,\\
 p ^{\frac{1-2\nu(\ga-1 )}{\ga \nu}} & \mbox{for}\ \ 1<\ga<\frac{3}{2}\ ,\\
 p^{\frac{1- 2\nu (\ga-1) }{\ga\nu }} & \mbox{for}\ \ \frac{3}{2}<\ga<2\ .
\end{cases}
\ee
These are non-divergent when $0<\ga<1$ as well as when $ 1<\ga<2$ and $0 < \nu < 1/(2\ga-2)$. Accordingly, if $0< \nu<1/2$, these quantities are regular for a generic $\ga\in (0, 2)$. Note that in these models, $f$ is not analytic at $\calR=0$. This example is just to show that if one relaxes the hypothesis of analyticity, it is possible to regularly match the inside and outside solutions, for a value of the polytropic parameter $\ga \in (0, 2)$.


\section{Swiss-Cheese: matching with FLRW}
\label{sec:cheese}

When unphysical situations are pinpointed \citep[see][]{Olmo}, one could try to build a mathematical model in which they do not appear. We examine such a mathematical representation, where the density $\rho$ could not reach zero, as its minimum value is given by the mean density of the Universe, namely $\rho_{\mathrm{u}}\approx 10^{-30}\,[\mathrm{g}/\mathrm{cm}^{3}]$.  
The simplest cosmological model of spherically symmetric gravitationally bound systems, embedded in an expanding Universe, is the GR regime described by Einstein-Straus model of Universe, also called the \emph{Swiss-cheese} model \citep[see][]{swisscheese}. 
Here, a hole is created by removing the background material inside a spherical boundary and {assuming all the mass is concentrated in the centre of the sphere. Mathematically, this representation is composed of a Schwarzschild metric smoothly connected to a Friedmann Universe. This translates into the
matching across a spherical boundary of a FLRW metric (as exterior solution) to a Schwarzschild metric (as interior solution). The spherical boundary stays at a fixed coordinate radius in the FLRW frame, but it changes with time in the Schwarzschild system \citep[see][]{Dyer}.\\
We now consider two different coordinate systems outside and inside: $[\tau,\si,\theta,\phi]$ and $[t,r,\theta,\phi]$, respectively.\\
The general FLRW metric for the external solution can be written in spherical coordinates as
\be 	
\label{eq:19}
g_{F}= -d\tau^{2}+a^{2}(\tau)\left[\frac{d\si^{2}}{1-K\si^{2}}+\si^{2}(d\theta^{2}+\sin^{2}\theta d\phi^{2})\right]\ ,
\ee
where $a(\tau)$ is the scale factor and $K = 0,\ \pm1$ is the constant curvature of space.\\
By imposing the Darmois junction conditions on a spherical hypersurface $\Sigma$, we try to match smoothly the metric (\ref{eq:19}) to our ansatz
for the inner region, which replaces the inner Schwarzschild metric in the standard Swiss-cheese model.\\
The matching hypersurface $\Sigma$  in the outer region will have equation $\si=\si_0$ and will be parametrized by coordinates $k^A=(u, \theta,\phi)$.
The canonical embedding of $\Sigma$ in the outer region is given by
\be
i:\Sigma\arr M: (u, \theta, \phi)\mapsto (\tau(u), \si_0, \theta, \phi)\ . 
\ee
The same hypersurface $\Sigma$ is also embedded in the inner region as
\be
j:\Sigma\arr M: (u, \theta, \phi)\mapsto (t(u), r(u), \theta, \phi)\ . 
\ee
By standard argument, one can define the first and second fundamental forms on the hypersurface $\Sigma$ with respect to each of the embeddings.
The first and second fundamental forms with respect to the embedding $i$ in the outer region will be denoted by
\be
\mathbb{I}^{o}= [\mathbb{I}^{o}]_{AB} dk^A\otimes dk^B\ ,
\qquad\qquad
\mathbb{II}^{o}= [\mathbb{II}^{o}]_{AB} dk^A\otimes dk^B\ ,
\ee
respectively, while the first and second fundamental forms with respect to the embedding $j$ in the inner region will be denoted by
\be
\mathbb{I}^{i}= [\mathbb{I}^{i}]_{AB} dk^A\otimes dk^B\ ,
\qquad\qquad
\mathbb{II}^{i}= [\mathbb{II}^{i}]_{AB} dk^A\otimes dk^B\ ,
\ee
respectively.

The non-null components of the two fundamental forms of $\Sigma$ given in an implicit form, i.e. $\si=\si_{0}$, with respect to the external  
space-time, are  
\be
\begin{cases}
[\mathbb{I}^{o}]_{11}=-\(\tau'\)^2\ ,\\
[\mathbb{I}^{o}]_{22}=a^{2}\ \si^{2}_{0}\ ,\\
[\mathbb{I}^{o}]_{33}=a^{2}\ \si^{2}_{0}\ \sin^{2}\theta\ ,\\
\end{cases}
\qquad
\begin{cases}
[\mathbb{II}^{o}]_{11}=0\ ,\\ 
[\mathbb{II}^{o}]_{22}=a\ \si_{0}\ \zeta\ ,\\
[\mathbb{II}^{o}]_{33}=a\ \si_{0}\zeta\ \sin^{2}\theta\ ,
\end{cases}
\ee
where we have set $\zeta\equiv\sqrt{1-K\si^{2}_{0}}\neq0$. 
Let us remark that both the first and second fundamental forms are diagonal. 

As regards the inner solution, instead, we calculate the fundamental forms from the line element (\ref{InSol}). In this case, the hypersurface $\Sigma$ is given by $t=t(u)$, $r=r(u)$, and we set $A=A\left[r(u)\right]$, $B=B\left[r(u)\right]$. For the non-zero components we obtain 
\be
\begin{cases}
[\mathbb{I}^{i}]_{11}=(r')^2 B-(t')^2 A\ ,\\ 
[\mathbb{I}^{i}]_{22}=r^{2} \ ,\\ 
[\mathbb{I}^{i}]_{33}=r^{2}\sin^{2}\theta\ ,
\end{cases}
\ee
\be
\begin{cases}
[\mathbb{II}^{i}]_{11}= -\frac{1}{2} \{ t' \left[ \Phi^2-(r')^2 B \right])A'
+ t'  (r')^2 AB'+2AB(t'r''-r't'')\} \frac{\Phi^{-1}}{\sqrt{AB}}\ ,\\ 
[\mathbb{II}^{i}]_{22}=rt' \sqrt{\frac{A}{B}} \Phi^{-1}\ ,\\
[\mathbb{II}^{i}]_{33}=rt' \sqrt{\frac{A}{B}}  \Phi^{-1}\sin^{2}\theta\ ,
\end{cases}
\ee
where we have set $\Phi^2=(t')^2 A-(r')^2 B$. Here, $A'$ and $B'$ are the derivatives with respect to the coordinate $r$
(then evaluated at $r=r(u)$), while $r'$ and $t'$ ($r''$ and $t''$) denote derivatives with respect to the parameter $u$, along the cutting hypersurface.\\
Thus, we examine the matching conditions between the FLRW metric of equation (\ref{eq:19}) outside and the metric of equation (\ref{InSol}) inside. The condition $[\mathbb{I}^{o}]_{ij}=[\mathbb{I}^{i}]_{ij}$ implies
\be\label{eq:20}
\left\{
\begin{aligned}
& (t')^2 A-(r')^2 B = (\tau')^2\ ,\\
& a^{2}\si^{2}_{0}\ =\ r^{2}\, 
\end{aligned}
\right.
\ee
On the other hand, the condition $[\mathbb{II}^{o}]_{22}=[\mathbb{II}^{i}]_{22}$ translates into the following equation
\be
a\ \si_{0}\ \zeta\  = rt' \sqrt{\frac{A}{B}} \Phi^{-1}\ ,
\ee
This, together with equations (\ref{eq:20}), can be recast as
\be
 (t')^2  = -\frac{\zeta^2B^2}{A(1-\zeta^2B)}(r')^2\ ,		
\qquad\qquad (\tau')^2  = - \frac{B}{(1-\zeta^2B)}(r')^2\ .
\ee
Accordingly, one cannot choose $r=\hbox{\it const}$ since it would imply $r'=0$ and $t'=\tau'=0$, i.e.~the matching surface $\Sigma$ would degenerate to a $2$-surface. By using the expression for $(t')^2$, we obtain
\be
(t')^2r''-r't't''=\zeta^2B\,\frac{A(2-\zeta^2 B) B'- B(1-\zeta^2 B)A'}{2 A^2(1-\zeta^2B)^2}\, (r'^4)\ .
\ee

Finally, we need to check $[\mathbb{II}^{o}]_{11}=[\mathbb{II}^{i}]_{11}$, which can be written as
\be
\label{eq:check}
(r')^2 \,\frac{B^2\zeta^2(AB)'}{A(1-B\zeta^2)^2}\, = \, 0 \ .
\ee
Since one cannot take $r'=0$, the other possibility is to have $(AB)'=0$. Under this assumption, equation (\ref{eq:check}) is identically satisfied and the matching conditions read as
\be	
\label{eq:22}
\begin{cases}
 (\tau')^2=  - \frac{B}{(1-\zeta^2B)}(r')^2\ , \\
 (t')^2=  -\frac{\zeta^2B^2}{A(1-\zeta^2B)}(r')^2\ ,\\
 a^{2}\si^{2}_{0}\ =\ r^{2}\ .\\
\end{cases}
\ee

{\small 
The original Swiss cheese model is recovered by setting the inner solution as Schwarzschild (and consequently $AB=1$) and fixing $\tau'=1$ as ansatz. Under these assumptions, one obtains
\be
\begin{cases}
(r')^2= \zeta^2 - A =   \zeta^2 - 1+ \frac{2M}{r}\ ,\\
 (t')^2=  \frac{\zeta^2r^2}{(r- 2M)^2}\ , \\
 a^{2}\si^{2}_{0}\ =\ r^{2}\ .\\
\end{cases}
\ee
The first two equations reveal us information about the cutting procedure for any provided initial conditions $(t_0, r_0)$. The third condition, instead, constrains $(t_0, r_0)$ and has to be preserved during the evolution. In fact, by differentiation one has 
$$
\dot a\si_0 = r'= \sqrt{\zeta^2 - 1+ \frac{2M}{r}} \ \ \ \ \ \ \ \ \ \ \frac{\dot a^2+K}{a^2} =  \frac{2M}{r^3}  \propto \rho\ .
$$
Consequently, having
\be
r''=-\frac{M}{r^2} \ \ \mbox{and}\ \ \frac{\ddot a}{a} =  -\frac{M}{ r^3}\ ,
\ee
one gets
\be
\frac{2\ddot a}{a} + \frac{\dot a^2+K}{a^2} = 0 = p 
\ee
Along the matching surface the pressure is continuous and zero, while the density is positive. Hence, the Schwarzschild metric can be matched with a FLRW solution.
}

It is quite straightforward to recover the usual polytropic matching in standard GR. In this case one matches an internal (non-vacuum) solution of Einstein equations with an external Schwarzschild solution. The matching surface is chosen to be $r=r_0$ and the matching of the second component of the first fundamental form implies that the radial coordinate is the same in both the interior and the exterior. The time coordinate $t$ is also set to be the same inside and outside. Thus, the matching conditions read as:
\[
B^{(o)}=B^{(i)}\ ,\ \  A^{(o)}=A^{(i)}\ , \ \  \left(A^{(o)}\right)'=\left(A^{(i)}\right)'\ ,
\]
which are verified for standard GR (see Appendix), in the special case of $f'=1$, $f''=0$. No divergence arises, since in GR $C=1$ and, since $f'=1$, $C_r\equiv 0$.
In general, there are two equations for determining the surfaces to cut and match. Initial conditions are given so that
$a^{2}\si^{2}_{0}\ =\ r^{2}$ at $u=0$.  Therefore, the matching is continuous if and only if the previous condition is preserved along evolution of $\left[t(u), r(u)\right]$ and $(AB)'= 0$ along the matching surface.
The real difficulty is the matching of the pressure on the surface, namely the matching of the $p(u)$ value given by the exterior model with the inner value.The exterior dynamics depends on the function $f(\calR)$ and on an external EoS.
The function $A(r)$ and $B(r)$ are determined by the internal equations and one can require initial conditions such that $AB=const$. Whether this is preserved by the inner dynamics and is valid for all $u$ is not clear in general. Further investigations should be devoted to study if it is possible to make a toy model in which the matching can be done and singularities are avoided. However, it is evident that the matching should be done in general by connecting the first and second fundamental forms. 
In \cite{no-go} it has been shown that there is no smooth matching if one assumes the surface $\Sigma$ to be in the form $r=r_0$, which is however a subclass of possible matchings.

\section{Conformal transformations}
\label{sec:conf_trans}

We consider now the behaviour of a polytropic matching under conformal transformations. We decide to use $\tilde g$ which obeys Einstein-like field equations, differently from $g$. In \cite{NostroOlmo} it was shown that for a fluid energy momentum tensor $T_{\mu\nu}$ there is a fluid effective energy-momentum tensor
\be
\tilde \T_{\mu\nu}=(\tilde \rho + \tilde p)\tilde u_\mu \tilde u_\nu + \tilde p \tilde g_{\mu\nu}\ ,
\ee
where the new effective pressure and density are given by
\be
\tilde \rho =  \frac{\rho}{\phi^2} +\frac{\calR(T)\phi + \ka^2 T}{4\ka^2 \phi^2}\ ,\ \ \tilde p =  \frac{p}{\phi^2} -\frac{\calR(T)\phi +\ka^2 T}{4\ka^2 \phi^2}\ .
\ee
By assuming $f(\calR)= \calR + \frac{\ep}{2}\calR^2$ (as in {Section \ref{re-exam}), we obtain
\be
\tilde \rho =  \frac{4\rho+\ep\ka^2 T^2 }{4(1-\ep\ka^2 T)^2}\ ,\ \ \ \ \tilde p=  \frac{4p-\ep\ka^2 T^2}{4(1-\ep\ka^2 T)^2}\ .
\ee
 
If the original metric $g$ is spherically symmetric, then one expects the conformal factor $\phi$ to depend only on the radial coordinate, the conformal metric $\tilde g$ being also spherically symmetric. Namely, we have
\be
g=-A(r)dt^2 + B(r)dr^2 + r^2d\Om^2
\iff
\tilde g=-\tilde A(\tilde r)dt^2 + \tilde B(\tilde r)d\tilde r^2 + \tilde r^2d\Om^2\ ,
\ee
where we have rescaled the radial coordinate $\tilde r^2 = \phi r^2$ and set
\be
\tilde A =  \phi A\ , \ \ \ \ \tilde B =  \frac{4 \phi^3 r^4}{(\phi' r^2+ 2\phi r)^2}B\ .
\ee
If we now introduce a pEoS on $(\tilde \rho, \tilde p)$, we would recover the GR case, where the matching of an inner solution with a Schwarzschild external solution $\tilde g$ is possible. On the other hand, if we impose a pEoS on $(\rho, p)$ (as done in proving the no-go theorem), we can simplify for small $p$ as follows
\be \label{pressureEoS}
{p =}
\begin{cases}
(\ga-1)\, \rho & \mbox{for}\ \ 0<\ga<1 \ ,\\ 
K\, \rho^\ga & \mbox{for}\ \ 1<\ga<2\ .
\end{cases}
\ee
In this case, the conformal factor reads as
\be
\phi= f'(\calR(T))=1 - \ep\ka^2 (3p-\rho) \sim
\begin{cases}
 1- \ep\ka^2 (3\ga-4)\rho & \mbox{for}\ \ 0<\ga<1\ , \\
 1- \ep\ka^2 \rho  & \mbox{for}\ \ 1<\ga<2\ .
\end{cases}
\ee
For $0<\ga<1$, hence $T\sim (3\ga-4)\rho$ and, in view of equation (\ref{pressureEoS}), for small $p$ the following hold
\be
\tilde \rho\sim   \frac{4\rho+\ep\ka^2  (3\ga-4)^2\rho^2 }{4(1-\ep\ka^2  (3\ga-4)\rho)^2}\sim \rho\ ,\ \ \ \ \tilde p\sim  \frac{4(\ga-1)\rho-\ep\ka^2  (3\ga-4)^2\rho^2}{4(1-\ep\ka^2  (3\ga-4)\rho)^2}\sim  (\ga-1)\rho\ ,
\ee
implying that $\tilde p \sim (\ga-1)\tilde \rho$.\\
For $1<\ga<2$, instead, $T\sim 3K \rho^\ga-\rho\sim -\rho$ and for small $p$ we can simplify as follows
\be
\tilde \rho =  \frac{4\rho+\ep\ka^2 \rho^2 }{4(1+\ep\ka^2 \rho)^2}\sim \rho\ , \ \ \ \ \tilde p =  \frac{4K\rho^\ga-\ep\ka^2 \rho^2}{4(1+\ep\ka^2 \rho)^2} \sim K\rho^\ga\ ,
\ee
implying that $\tilde p \sim K\tilde \rho^\ga$.\\
Note that an exact pEoS for $(\rho, p)$ implies a varying pEoS for $(\tilde \rho, \tilde p)$. However, pEoS is preserved by conformal transformations, at small $p$. To summarize, conformal transformations map Palatini $f(\calR)$-theories for $g$ into an effective standard GR for $\tilde g$.
The spherically symmetric ansatz, the polytropic equation of state and the form of source energy-momentum tensor are preserved. In the standard formulation for $\tilde g$, polytropic stars are easily produced, thus it should be possible to map $\tilde g$ back to the original framework for $g$. However, the no-go theorem shows that there might be no matching available for $g$. The only possible explanation for this is that the conformal transformation itself is singular at the matching surface. The conformal factor in the interior reads as
\be
\phi =
\begin{cases}
1- \ep\ka^2 (3\ga-4)\tilde \rho  & \mbox{for}\ \ 0<\ga<1 \ ,\\
1- \ep\ka^2 \tilde \rho  & \mbox{for}\ \ 1<\ga<2\ ,\\
\end{cases}
\ee
while $\phi= 1$ in the exterior. 
The conformal factor is thence continuous, though not differentiable at the surface. This means that a $\calC^1$ matching for $\tilde g$ is mapped into a $\calC^0$ matching for $g$. In fact, 
the second fundamental forms for the matching surface with respect to $g$ and $\tilde g$ are related by
\be
{{\tilde{[\mathbb{II}}^{o}]}_{AB}}= \sqrt{\phi} {[\mathbb{II}^{o}]_{AB}} +\frac{1}{2}\( g^{\al\ep} g_{\mu\nu}-2\de^\al_{(\mu}\de^\ep_{\nu)}\)u_\al \na_\ep \ln\phi \> {J^\mu_AJ^\nu_B}\ ,
\ee
where $J^\mu_A$ denotes the Jacobian of the embedding of the matching surface into $M$.
One can see that the transformation law 
depends on $\phi'$ which is discontinuos at the matching surface. 
Let us denote by ${{\tilde {[\mathbb{II}}^{o}]}}$ and  ${{[\mathbb{II}^{o}]}}$ the matrices with entries ${{\tilde {[\mathbb{II}}^{o}]}}_{AB}$ and ${{[\mathbb{II}^{o}]}}_{AB}$, respectively, 
then the difference between the two matches is measured as 
\be
{{\tilde {\mathbb{II}}^{o}}}- \sqrt{\phi} {{\mathbb{II}^{o}}} = 
\frac{1}{\Phi\sqrt{AB}}\frac{t' \phi'}{2\phi}
\(\begin{matrix}
A \Phi^2&0&0\\
0&- r^2 A &0\\
0&0&- r^2 A\sin^2(\theta) \\
\end{matrix}\)\ ,
\ee
which cannot vanish along the matching surface. To conclude, even when both the matchings are possible, matching $\calC^1$ with respect to $g$ is not the same as matching $\calC^1$ with respect to $\tilde g$. The choice of the correct metric to be matched} is a matter of interpretation of the model and its physical meaning.

\section{Conclusion and perspectives}
\label{sec:concl}

We showed that in Palatini $f(\calR)$-theories problems might arise in matching inner and exterior solutions for stars, when using a pEoS and for analytic $f(\calR)$. In particular, for some values of the polytropic index $\left(3/2<\ga<2\right)$ and a generic analytic $f(\calR)$, a singularity develops near the surface of the star. We also proved that matching the conformal metric $\tilde g$ is possible as in standard GR. In this case, the singularity is in the conformal transformation, which is $\calC^0$ instead of $\calC^2$, affecting the matching of the second fundamental forms.\\
Naturally, the interpretation of the theory gives a hint on which is the {\it physical metric} involved. If one assumes the free fall of particles being 
governed by $g$, then the theory is equivalent to Brans-Dicke theory with a potential and $\om=-3/2$ \citep[see][]{Kepler}.
However, this value is ruled out by Solar System tests (though assuming no potential). In order to proceed under these assumptions, the effect of the potential has to be also considered.\\
On the other hand, we believe that there exist good motivations for assuming that the free fall of particles is ruled by $\tilde g$. 
In fact, by universality theorem \citep[see][]{Universality} vacuum solutions are described by metrics which are also solutions of
GR with a (suitable) cosmological constant depending on the function $f(\calR)$.
If we are modelling for example the Solar System, then we have one of these solutions and, since we know that the cosmological constant has no effect at these scales, we can reasonably assume that $f(\calR)$ should be such that it corresponds to a cosmological constant which is small enough to agree with Solar System experiments. There are plenty of models for which the cosmological constant has no effect at small scales though it becomes relevant at bigger scales.

Beside this motivation, Ehlers-Pirani-Schild (EPS) proposed an interpretation of gravitational theories based on potentially observable quantities (worldlines of massive particles and light rays). They showed that the geometry of space-time is described by a conformal class of metrics $[g]$ and a projective class of connections $[\tilde \Ga]$. In this framework, the free fall is associated by construction to the connection $\tilde \Ga$ and it suggests a Palatini approach \citep[see][]{EPS,EPS1,NostroOlmo}. Moreover, Palatini $f(\calR)$-theories are compatible to EPS formalism \citep[see][]{EPS2}. Within the EPS interpretation of Palatini $f(\calR)$-theories, it is natural to search for matching of $\tilde g$ rather than matching of $g$: $\tilde g$ is in fact the {\it physical} metric because it describes the free fall. In particular, this is worth to use when polytropic stars are considered. In fact, even when a matching of $g$ is possible, the two matchings are not equivalent and we suggest the matching of $\tilde g$ is physically more meaningful.

Let us remark that we did not imposed anywhere energy conditions, neither on $\T_{\mu\nu}$ or on $\tilde\T_{\mu\nu}$, and we did not declared what is the physical meaning of the original metric $g_{\mu\nu}$. The reason not to do that is that nothing in this paper depends on these assumptions.
However, in view of EPS framework one should assume that the metric $g$ is related to operational definition of distances (since choosing a representative in the conformal structure is in 1--to--1 correspondence with choosing a definition for distances and time lapses). 
One should also investigate if ordinary quantum mechanics is the one for inertial observers with respect to $g$ or $\tilde g$.
The differences between the two metrics are probably too tiny to be tested experimentally and consequently also this assumption is an independent choice,
 which does not affect the content of this paper.
 It is our opinion that, since our protocols for distances strongly rely on quantum mechanics, the frame for quantum mechanics should be selected to be $g$ as well. 
 
However, what is relevant here is that that opens a new perspective to be explored in the long standing issue of which is the {\it physical frame}.
In the past studies, it was often, if not always, assumed that in the end one of the frames would have eventually emerged as {\it the} physical frame and 
one would have done everything with one metric (free fall, distances, quantum mechanics, minimal coupling, energy conditions, \dots) while the other metric(s)
would appear as auxiliary object(s) with no specific physical meaning.
Let us stress that there is another option, and it needs to be investigated, namely the possibility that different frames appear in the theory since they have different meaning and one could have a reasonable model in which free fall, energy conditions, minimal coupling, quantum mechanics are done in different frames.
This possibility corresponds to effects which would be characteristic of extended gravity theories which have been almost entirely overviewed. 
This would give different frames a new and different physical meaning.
 
In future investigation, one should also better examine whether it is possible to build a model by matching the inner solution to a FLRW solution, where the density does not approach zero and no singularities appear, as suggested by \cite{Olmo}. Furthermore, one could show whether non-analytic models can avoid singularities, as suggested by our analysis.

\section*{Appendix A: Inner solution in dimension $m=4$}
\label{App4}

We hereafter examine Einstein field equations for the inner solution, in dimension $m=4$ \citep[see][]{no-go}. The independent equations are obtained by plugging the metric ansatz in equation (\ref{InSol}) into equation (\ref{fe}) and read as follows:
\be
\begin{aligned}
&A \frac{B' r+B^2-B}{r^{2}B^2}  = \frac{ \kappa^{2}}{f'}  A\rho
+ \frac{1}{2}A\left(\calR-\frac{f}{ f'}\right)+ \frac{A}{B}\left[ \frac{\partial_{rr}f'}{f'}  %
  +\left(\frac{2}{r} -\frac{B'}{2B} \right)\frac{\partial_rf'}{f'} 
  - \frac{3}{4}\left(\frac{\partial_rf' }{f'}\right)^{2}\right]\ ,\\
&\frac{A' r-AB+A}{r^{2}A}  =\frac{ \kappa^{2}}{f'} B p
-\frac{1}{2}B\left(\calR-\frac{f}{ f'}\right)
-\left[\left(\frac{A'}{2A}+\frac{2}{r} \right)\frac{\partial_rf' }{f'}
  +\frac{3}{4}\left(\frac{\partial_rf'}{f'} \right)^{2}\right]\ ,\\
& \frac{2A'AB-2B'A^2+ 2rA''AB-rB(A')^2 -rAA'B'}{4A^2B^2}\, r = \\
& = \frac{ \kappa^{2}}{f'} r^2 p
 -\frac{1}{2}r^2\left(\calR-\frac{f}{ f'}\right)-\frac{r^2}{B}\left[\frac{\partial_{rr}f' }{f'}
 +\left(\frac{1}{r}-\frac{B'}{2B} + \frac{A'}{2A}\right)\frac{\partial_rf' }{f'}
 -\frac{3}{4}\left(\frac{\partial_rf' }{f'}\right)^2\right]\ .\\
\end{aligned}
\label{FieldEqs}
\ee
Here, primes on $A$,$B$ and $f$ denote derivatives with respect to their argument, namely $r$ and $\calR$, respectively. Note that there is another equation associated to the angular coordinate $\phi$, which is equivalent to the equation associated to the coordinate $\theta$. The first two field equations can be recast in the following form:
\be
\label{eq:5}
\begin{aligned}
\frac{A'}{A}=&-\frac{1}{1+\sigma}\left(\frac{1-B}{r}-\frac{B}{f'}\kappa^{2} r p +\frac{\alpha}{r}\right),\ \  \\
\frac{B'}{B}=&\frac{1}{1+\sigma}\left(\frac{1-B}{r}+\frac{B}{f'}\kappa^{2} r \rho +\frac{\alpha+\beta}{r}\right),
\end{aligned}
\ee
where we set 
\be\label{eq:6}
\begin{aligned}
&\alpha \equiv r^{2} \left[\frac{3}{4}\left(\frac{\partial_r f'}{f'}\right)^{2}+\frac{2\partial_rf'}{rf'}+\frac{B}{2}\left(\calR-\frac{f}{f'}\right)\right],\\
& \beta \equiv r^{2} \left[\frac{\partial_{rr}f'}{f'}-\frac{3}{2}\left(\frac{\partial_rf'}{f'}\right)^{2}\right],
\qquad\qquad
 \sigma \equiv \frac{r\partial_rf'}{2f'}\ .
\end{aligned}
\ee
The conservation law for the energy-momentum tensor, i.e. $\nabla_\nu\T^{\mu\nu}=0$, 
which is a consequence of Bianchi identities in equation (\ref{fe}), can be expressed as
\be
\label{eq:7}
p_r {=} \frac{-1}{1+\sigma} \frac{(\rho+p)}{r[r-2M(r)]}\left\{M(r)+\frac{\kappa^{2}r^{3} p}{2f'}-\frac{\alpha}{2}\left[r-2M(r)\right]\right\}\ , 
\ee
where we set $p_r{\equiv}p'(r)$ for the derivative of the pressure with respect to the coordinate $r$ and
\be\label{eqM}
M(r){\equiv} \frac{r\,(B-1)}{2B}\ ,
\ee
from which it follows that
\be
B = \frac{r}{r-2M(r)}\ .
\ee

By fixing an EoS $\rho = \rho(p)$ for the matter fluid, one has three unknown functions of $r$, namely $(p, A, B )$. Equations (\ref{eq:7}) and (\ref{eq:5}) are an ODS in normal form, which in principle can be solved. Although the system is generally determined, the solution is in practice hard to find in an explicit form. In view of (\ref{eq:5}), we transform equation (\ref{eqM}) into a differential equation for $M(r)$, as 
\be\label{eq:8}
M'(r)=\frac{1}{1+\sigma} \left[\frac{\kappa^{2}r^{2}\rho}{2f'}+\frac{\alpha+\beta}{2}-\frac{M(r)}{r}\left(\alpha+\beta-\sigma\right)\right]\ .
\ee
It follows that the third field equation in (\ref{FieldEqs}) is identically satisfied.\\

If we now consider a pEoS for the fluid
\be 
\label{eq:EoS}
\rho= \left(\frac{p}{K}\right)^{\frac{1}{\gamma}} + \frac{p}{\gamma-1}\ ,
\ee
then it follows that
\be
\frac{\mathrm{d}p}{\mathrm{d}\rho}=\frac{\ga K^{\frac{1}{\ga}}}{  p^{\frac{1-\ga}{\ga}}+ \frac{\ga}{\ga-1}  K^{\frac{1}{\ga}} }\ .
\ee
Here $\gamma$ and $K$ are two real constants which depend on the fluid: in particular $\gamma\in (0, 1)\cup(1,2)$ and $K>0$. Notice that $T = 3p - \rho$ and
\be \label{eq:12}
f'_\rho 
{\equiv \frac{\mathrm{d}f'}{\mathrm{d}\rho} } 
=\frac{\kappa^2 f''}{  f''\calR -f'} \left(\frac{3\ga K^{\frac{1}{\ga}}}{p^{\frac{1-\ga}{\ga}} + \frac{\ga K^{\frac{1}{\ga}}}{\ga-1}} -1\right)
\ ,
\ee
This stays finite while approaching the stellar surface (as $p$ tends to zero) and has a limit which depends on the parameter $\ga$, namely
\be
f'_\rho \arr
\label{eq:12b}
\begin{cases}
\frac{\kappa^2 f''}{  f''\calR -f'} \left(3\ga-4\right) & \mbox{for}\ \ 0<\ga<1 \ ,\\
- \frac{\kappa^2 f''}{  f''\calR -f'} & \mbox{for}\ \ 1<\ga<2 \ .
\end{cases}
\ee
Note that for a generic analytic $f$, this quantity is finite. In fact, the denominator is computed at a zero of the vacuum master equation and it is non-vanishing if the zero is simple, which is generally true. Note also that in standard GR $f''=0$ and $f'_\rho=0$, for any value of polytropic parameters. Furthermore, we can compute the quantity $f'_{\rho\rho} {\equiv} \frac{d^2f'}{d\rho^2}$ as
\be
\label{eq:12c}
f'_{\rho\rho} {\equiv} 
\frac{\kappa^4f'''}{\( f''\calR- f' \)^2}\ \left(3 \frac{\mathrm{d}p}{\mathrm{d}\rho}-1\right)^2
- \frac{\ka^4 f''' f''\calR }{ (f'' \calR -  f')^3}\ \left( 3 \frac{\mathrm{d}p}{\mathrm{d}\rho}-1\right)^2 
-\frac{3\kappa^2 f'' }{ f''\calR- f' }\ \frac{(1-\ga)   }{\ga \( p^{\frac{1-\ga}{\ga}}+ \frac{\ga K^{\frac{1}{\ga}}}{ \ga-1  }  \)} \(\frac{\mathrm{d}p}{\mathrm{d}\rho}\)^2 p^{\frac{1-2\ga}{\ga}}
\ee

The limit of $f'_{\rho\rho}$ to the stellar surface as $p$ tends to zero depends again on the value of $\ga$
\be 
\label{eq:12d}
f'_{\rho\rho} \arr
\begin{cases}
-\frac{\kappa^4f''' f' }{\( f''\calR- f' \)^3}\  \left(3 \ga-4\right)^2  & \mbox{for}\ \ 0<\ga<\frac{1}{2} \ ,\\
\frac{3\kappa^2 f'' }{ f''\calR- f' }\   \frac{(1-\ga)^4   }{ \ga^2 K^{\frac{1}{\ga}}  }  p^{\frac{1-2\ga}{\ga}} & \mbox{for}\ \ \frac{1}{2}<\ga<1 \ ,\\
-\frac{3\kappa^2 f'' }{ f''\calR- f' }\    \ga(1-\ga) K^{\frac{2}{\ga}}   p^{\frac{\ga-2 }{\ga}}  & \mbox{for}\ \ 1<\ga<2 \ ,\\
\end{cases}
\ee
though it does not stay always finite in this case. 
It follows that:
\be
(\rho+p)\frac{d\rho}{dp} = \left(\frac{p^{\frac{1}{\gamma}}}{K^{\frac{1}{\gamma}}} + \frac{ \gamma p }{\gamma-1}\right) \left(\frac{1}{\gamma}\frac{p^{\frac{1}{\gamma}-1} }{K^{\frac{1}{\gamma}} }+ \frac{1}{\gamma-1}\right) = \frac{1}{\gamma}\frac{p^{\frac{2}{\gamma}-1} }{K^{\frac{2}{\gamma}} }
+ \frac{ 2p^{\frac{1}{\gamma}}}{(\gamma-1) K^{\frac{1}{\gamma}}} + \frac{ \gamma p }{(\gamma-1)^2}\ ,
\ee
which approaches zero as $p$ tends to zero, if $0<\gamma<2$.

We multiply equation (\ref{eq:7}) by $f'_p{\equiv}\mathrm{d}f'/\mathrm{d}p$ and use equation (\ref{eq:6}) to get a quadratic expression in $\partial_ r f'$. Solutions of the latter are 
\be\label{eq:9}
\partial_ r f' =\frac{-4rf'(C-f')(r-2M)\pm\sqrt{2\Delta}}{r^{2}(3C-4f')(r-2M)}\ ,
\ee
where we set
\be\label{eq:10}
\begin{aligned}
&C{\equiv}\frac{\mathrm{d}f'}{\mathrm{d}T}\(3-\frac{\mathrm{d}\rho}{\mathrm{d}p}\)(\rho+p)= f'_\rho\> \frac{\mathrm{d}\rho}{\mathrm{d}p}(\rho+p)  \ ,\\
&\Delta{\equiv}f'r^{2}(r-2M)\left\{ 8f'(C-f')^{2}(r-2M)- C(4f'-3C)\left[\left(2\kappa^{2}p-f'\calR+f\right)r^{3}+4f'M\right]\right\}\ .
\end{aligned}
\ee
To summarize, when the stellar surface is approached from the inside (i.e. as $r$ tends to $r_0^-$), then $p$ tends to zero, so that also $\rho$ tends to zero. This implies that $T$ approaches zero as well, while $f$ tends to $f'\calR_0/2$, where $\calR_0$ is a solution of master equation in vacuum. Hence, $C$ approaches zero, with $\mathrm{d}f'/\mathrm{d}\rho$ being finite.} \\
We set  $M_0{\equiv} M(r_0)$ so that, in view of equation (\ref{eqM}), we can write $B|_{r_0}= \left(1-2 M_0/r_0\right)^{-1}$.
It follows that
\be
2\Delta|_{r_0}{\equiv}  16 r_0^{2}(r_0-2M_0)^2 (f')^{4}\ .
\ee
The quantity $C$ is a function of $p$ (since the scalar curvature $\calR$ is a function of $T=3p-\rho$ because of the master equation
and the pressure density $\rho$ can be eliminated by using the EoS).
Moreover, since the pressure is a function of $r$ on-shell, one can define the derivative of $C$ with respect to $r$, which will be denoted by $C_r$.
Thus, by setting $C_r=p_{r}\,\mathrm{d}C/\mathrm{d}p$, equation (\ref{eq:9}) gives either  
\be
\partial_ r f' |_{r_0}=-\frac{2 f' }{r_0}\ ,
\qquad 
\partial_{rr}f' |_{r_{0}}= \frac{(4-\calR_{0}r^{2}_{0})\,C_r}{8\left(r_{0}-2M_0\right)} +\frac{2f'}{r_0^2}\ , 
\label{eq:14a}
\ee
or
\be
\partial_ r f' |_{r_0}=0\ ,
\qquad 
\partial_{rr}f'|_{r_{0}}=\frac{(\calR_{0}r^{3}_{0}-8M_0)\,C_r}{8r_{0}\left(r_{0}-2M_0\right)}\ .
\label{eq:14}
\ee
In the case of solution (\ref{eq:14}), one also has
\be
\left.\frac{A'}{A}\right|_{r_0}=\frac{8 M_0-\calR_0 r_0^{3}}{4r_0\,(r_0-2 M_0)} ,
\qquad
\left.\frac{B'}{B}\right|_{r_0}= \frac{ (2f' +r_0C_r)(\calR_0 r_0^{3}- 8 M_0)}{8r_0f'\left(r_{0}-2M_0\right)},
\ee
and
\be \label{eq:15}
p'|_{r_0}=0\ ,
\qquad\qquad
M' |_{r_{0}}=\frac{2f'\calR_{0}r^{2}_{0}+(\calR_{0}r^{3}_{0}-8M_0)C_r}{16f'}\ .
\ee
%
%
%
The exterior solution ${\bf g}$ is obtained by setting $p=0$ (and consequently $\rho=0$). The master equation simplifies to $f'\calR -2 f=0$, which sets a constant value for $\calR= {\bf R}_0$ in a quantized set encoded by the function $f$. This in turn implies that $\partial_r f'=0$ and $\partial_{rr} f'=0$. Vacuum field equations are thence given by
\be
{\bf G}_{\mu\nu}= -\Lambda {\bf g}_{\mu\nu}\ ,\ \ \ {\mbox{where}}\ \ \Lambda{\equiv} \frac{1}{2}\left({\bf R}_0 - \frac{f}{f'}\right)=  \frac{1}{4}{\bf R}_0\ .
\ee 
It is then reasonable to assume for the exterior solution the well-known Schwarzschild(-AdS) solution with cosmological constant:
\be
{\bf g}=-{\bf A}(r)\mathrm{d}t^{2}+{\bf B}(r)\mathrm{d}r^{2}+r^{2}\mathrm{d}\Omega^{2}\ ,
\ee
{where}
\be
{\bf A}(r){\equiv} 1-\frac{2 {\bf m}}{r} - \frac{{\bf R}_0}{12}r^2 \ \ \ {\mbox{and}}\ \ \ {\bf B}(r){\equiv} \frac{1}{{\bf A}(r)}\ .
\ee 
The exterior solution evaluated at the surface, namely $r=r_{0}$, gives
\be 
\label{eq:13}
\begin{aligned}
\left.\frac{{\bf A}'}{{\bf A}}\right|_{r_0} & =  \frac{2\,(r_0^3{\bf R}_0-12 {\bf m}) }{r_0\,(r_0^3{\bf R}_0-12r_0+24 {\bf m} )  }\ ,\\
\left.\frac{{\bf B}'}{{\bf B}}\right|_{r_0} & = - \frac{2\,(r_0^3{\bf R}_0-12 {\bf m}) }{r_0\,(r_0^3{\bf R}_0-12r_0+24 {\bf m} )  } \ .
\end{aligned}
\ee
The exterior and interior solution need to match on the stellar surface. This can be obtained by imposing continuity at $r=r_0$ of the coefficients $B(r_0)={\bf B}(r_0)$ and $(A'/A)(r_0)=({\bf A'}/{\bf A})(r_0)$, thus by matching the constants
\be
\calR_0= {\bf R}_0\ \ \ \ \ \mbox{and}\ \ \ \ M_0= {\bf m}+\frac{{\bf R}_0 r_0^3}{24}\ .
\ee
As a consequence, $A'$, $p$, $p'$  and $\partial_r f'$ are also continuous. 
We can now consider the following
\be
\left.\frac{{\bf B}'}{{\bf B}}\right|_{r_0}-  \left.\frac{{ B}'}{{ B}}\right|_{r_0}=\frac{r_0^3 \calR_0-12 {\bf m}}{r_0^3 \calR_0-12r_0+24 {\bf m}}\frac{C_r}{f'}\ .
\ee
To obtain continuity, one needs either to fine tune the parameters (so that $r_0^3 \calR_0=12 {\bf m}$) or have $C_r/f'$ approaching zero as $r$ tends to $r_0$. If $f$ is assumed to be regular at $r_0$, that amounts to have that $C_r$ tends to zero as $r$ tends to $r_0$. Thus, $B'$, $M'$ and $\partial_{rr} f'$ are continuous if $C_r$ tends to zero. Notice however, that more generally, if $f$ diverges at $r_0$, then $C_r$ is allowed to diverge, provided that the ratio $C_r/f'$ tends to zero. We focus now on $C_r$ at the surface:
\be
C_r=\frac{dC}{dp}p_r= \left[\left(\rho+p\right) f'_{pp}+  \( \frac{d\rho}{dp}+1 \) f'_p\right]p'= \left[f'_{pp}\left(\frac{p^{\frac{1}{\gamma}}}{K^{\frac{1}{\gamma}}} + \frac{\gamma p}{\gamma-1}\right) + f'_p\left( \frac{ p^{\frac{1-\gamma}{\gamma}}}{\gamma K^{\frac{1}{\gamma}}}+ \frac{\gamma }{\gamma-1}\right)\right]p_r\ ,
\ee
where we used the pEoS. In addition, the following hold:
\be
\begin{aligned}
f'_{p} =&\frac{df'}{d\rho}\frac{d\rho}{dp} 
=\(\frac{  p^{\frac{1-\ga}{\ga}}}{\ga K^{\frac{1}{\ga}}}+    \frac{1  }{ \ga-1  }\)f'_\rho \ ,\\
f'_{pp} 
=&    \left[ \frac{  p^{2\frac{1-\ga}{\ga}}}{\ga^2 K^{\frac{2}{\ga}}} +2\frac{  p^{\frac{1-\ga}{\ga}}}{\ga(\ga-1) K^{\frac{1}{\ga}}}  +    \frac{1  }{( \ga-1)^2  } \right] f'_{\rho\rho} 
+  \frac{ (1-\ga) p^{\frac{1-2\ga}{\ga} }}{\ga^2 K^{\frac{1}{\ga}}}f'_\rho\ ,
\end{aligned}
\ee
where we set 
$f'_{pp}{\equiv} \mathrm{d}^{2}f'/\mathrm{d}p^{2}$ and $f'_{\rho\rho}{\equiv} \mathrm{d}^{2}f'/\mathrm{d}\rho^{2}$.
By using equation (\ref{eq:7}), one gets:
\be
C_r \sim \frac{\calR_0 r_0-8M_0}{8r_0(r_0-2M_0)} \frac{(\ga-1)\, \si_1(p) f'_\rho+ \si_2(p) f'_{\rho\rho}}{\ga^2(\ga-1)^4}\ ,
\ee 
where by $\sim$ we denote equality of dominant terms, as $p$ tends to zero. 
The terms $\si_1(p)$ and $\si_1(p)$ read as:
\be
\begin{aligned}
\si_1(p) & = -(\ga-2)(\ga-1)^3 K^{-\frac{3}{\ga} } p^{\frac{3-2\ga}{\ga} } -\ga(\ga-4)(\ga-1)^2 K^{-\frac{2}{\ga} } p^{\frac{2-\ga}{\ga} }+\ga^2 (\ga-1)(\ga+2)K^{-\frac{1}{\ga} }p^{\frac{1}{\ga} }+p\ga^4 \ ,\\
\si_2(p) & =(\ga-1)^4K^{-\frac{4}{\ga} }p^{\frac{4-2\ga}{\ga} } +4\ga(\ga-1)^3K^{-\frac{3}{\ga} }p^{\frac{3-\ga}{\ga} }+4\ga^2(\ga^2-1)K^{-\frac{1}{\ga} }p^{\frac{1+\ga}{\ga} } +6\ga^2(\ga-1)^2K^{-\frac{2}{\ga} }p^{\frac{2}{\ga} }+ p^2\ga^4\ .
\end{aligned}
\ee 
Note that, referring to equation
(\ref{eq:12c}), we obtain that $f'_{\rho\rho}\si_2(p)$ tends to zero for generic vales of $\ga$ in the interval $(0,2)$.
As we showed that $f'_\rho$ is finite, for $0<\gamma< 3/2$ $C_r$ tends to zero. However, for $3/2<\gamma<2$, $C_r$ diverges and reads as
\be
C_r
\sim\frac{\calR_0 r_0-8M_0}{8r_0(r_0-2M_0)}\frac{2-\gamma}{\gamma^2K^{\frac{3}{\gamma}}}  p^{\frac{3-2\gamma}{\gamma}} f'_\rho\ .
\ee
The Ricci scalar of the metric $g$ at the surface is given by
\be 
\label{eq:18}
R = \frac{r(rA'+4A)(BA'-AB')+2Br^{2} (AA''- A'^2)-4A^2B^2+4BA^2}{2r^{2}A^2B^2}\ .
\ee
Furthermore, we obtain that at the surface
\be
R \arr 3 \frac{r_0^3\calR_0-8M_0}{8r_0^2}\frac{C_r}{f' }+\calR_0\ .
\label{DivergentR}
\ee
This means that, for $3/2<\gamma<2$ and regular $f$, $C_r/f'$ is divergent as the scalar curvature. Hence, the singularity at the surface is not a coordinate singularity. 
{We checked that also in the case of dimension $m=3$, the singularity at the surface is not a coordinate singularity.}

\section*{Acknowledgements}

We acknowledge the contribution of INFN (Iniziativa Specifica QGSKY), the local research project {\it  Metodi Geometrici in Fisica Matematica e Applicazioni (2014)} of Dipartimento di Matematica of University of Torino (Italy). This paper is also supported by INdAM-GNFM.

\label{lastpage}

\end{document}